\documentclass[conference]{IEEEtran}

\usepackage{amsmath,amssymb,amsfonts}
\usepackage{graphicx}
\usepackage{booktabs}
\usepackage{xcolor}
\usepackage[hidelinks]{hyperref}
\usepackage{cite}

\newcommand{\ket}[1]{\lvert #1 \rangle}
\newcommand{\zerol}{\ket{0}_L}
\newcommand{\plusl}{\ket{+}_L}
\newcommand{\bigo}[1]{\mathcal{O}(#1)}
\begin{document}

\title{Fast Unitary Preparation of Surface-Code Logical States on Neutral-Atom Hardware}

\author{%
\IEEEauthorblockN{Ludwig Schmid\textsuperscript{1,\,$\ast$} and Robert Wille\textsuperscript{1,2}}
\IEEEauthorblockA{\textsuperscript{1}Chair for Design Automation, Technical University of Munich, Munich, Germany\\
\textsuperscript{2}MQSC, Garching near Munich, Germany\\
\{ludwig.s.schmid, robert.wille\}@tum.de \quad \textsuperscript{$\ast$}\,Corresponding author}%
\vspace{-12pt}%
}

\maketitle

\begin{abstract}
Preparing logical states is a constantly recurring primitive at the start of
any surface-code-based quantum computation. The standard measurement-based
protocol is expensive on neutral atoms, since measurements are orders of
magnitude slower than gates and, on single-zone architectures, additionally
require shuttling the atoms to a readout zone. In this work, we
present a measurement-free, unitary preparation of surface-code Pauli eigenstates
tailored to neutral atoms. A \emph{bidirectional} stabilizer-expanding
CNOT cascade grows the patch from its middle line outward in depth
$(d{+}3)/2$, while retaining fault distance $d$ in the protected error
direction. Each layer maps to one collective atom move
followed by a single global Rydberg pulse, with about one row pickup per two
layers. We compile and simulate the resulting schedules with the open-source
\texttt{bloqade} toolchain under a hardware-calibrated, circuit-level
neutral-atom noise model. The bidirectional cascade achieves the lowest
logical error rate at every distance and is the only construction tested that
operates below threshold for both logical states at the considered
device-level noise.
\end{abstract}


\section{Introduction}\label{sec:intro}

Quantum error correction with surface codes has reached the sub-threshold
regime in experiment~\cite{acharya2024threshold,bluvstein2024logical}, moving
the cost of the underlying logical primitives into focus. Among these,
preparing fresh logical states is ubiquitous. The standard protocol
initializes all data qubits and runs ${\sim}d$ rounds of stabilizer
measurement. This is particularly expensive on neutral atoms, where
measurement is orders of magnitude slower than
gates~\cite{graham2023midcircuit,bluvstein2024logical} and, on single-zone
architectures, additionally requires shuttling the atoms to a readout
zone~\cite{bluvstein2024logical}.

Unitary, measurement-free preparation avoids this cost entirely. Instead of
measuring the stabilizers into their eigenstates, the idea is to \emph{grow}
them. A CNOT cascade expands the stabilizers of an initial product state
step by step to the full code state. A recent proposal showed that this can
be done fault-tolerantly against one error type at depth
$\bigo{d}$~\cite{ref:unitaryprep,zen2024rl}. Related unitary and
measurement-free encoders reach similar
depths~\cite{higgott2021optimal,tsai2025encoder,goto2023measfree}. These
constructions, however, are not optimized for neutral-atom machines. The
latter offer a specific, highly parallel operation set of
global Rydberg CZ layers~\cite{evered2023gates} and collective,
order-preserving row and column moves of atoms held in
acousto-optic-deflector (AOD) tweezers~\cite{bluvstein2022transport}, and a
preparation scheme matched to these capabilities has been missing.

In this work, we adapt and improve the unitary stabilizer-expanding
preparation for single-zone neutral-atom hardware. Our construction combines a middle-out
growth pattern, employed previously in a different context for the
preparation of topologically ordered states on a superconducting
processor~\cite{satzinger2021topo}, with the fault-tolerant expansion rule
of~\cite{ref:unitaryprep}. Growing the patch from its middle line in both
directions at once halves the depth and yields exactly the collective,
nearest-neighbour move structure that AOD shuttling favours. In summary, we
contribute:
\begin{itemize}
  \item a bidirectional middle-out cascade of depth $(d{+}3)/2=\bigo{d/2}$
        with unchanged protected-direction fault distance $d$ (compared to
        $d$ and $3(d{-}1)$~\cite{ref:unitaryprep});
  \item an AOD-native, single-zone realization with one collective
        nearest-neighbour move and one global Rydberg pulse per layer, and
        about one row pickup per two layers;
  \item a validation using the open-source \texttt{bloqade}
        toolchain~\cite{tool:bloqade} with a hardware-calibrated neutral-atom
        noise model, in which the bidirectional cascade beats both baselines
        and is the only construction below threshold for both logical states
        at the considered device-level noise.
\end{itemize}

\section{Background}\label{sec:background}

\subsection{Rotated surface code}\label{sec:background-code}
\begin{figure}[t]
  \centering
  \vspace{-15pt}%
  \includegraphics[width=0.64\columnwidth]{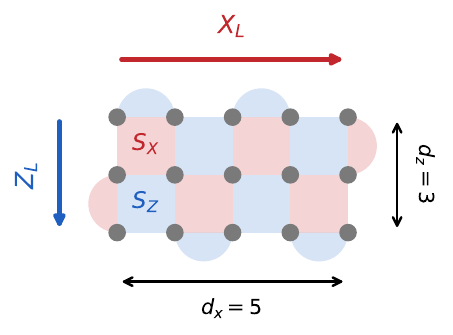}
  \vspace{-8pt}%
  \caption{Rotated surface-code patch ($d_x{=}5$ rows, $d_z{=}3$ columns):
  data qubits (grey), $X$-type (red) and $Z$-type (blue) stabilizers, and the
  logical strings $X_L$/$Z_L$ along the two directions.}
  \label{fig:patch}
\end{figure}

The surface code~\cite{bravyi1998boundary,dennis2002topological,fowler2012surface}
in its rotated layout~\cite{bombin2007rotated} places data qubits on a
$d_x\times d_z$ grid with $X$- and $Z$-type plaquette stabilizers arranged in a
checkerboard (Fig.~\ref{fig:patch}). The logical operators are strings of $X$
($X_L$) and $Z$ ($Z_L$) Paulis along the two grid directions. The code distance
against $Z$ errors is the number of columns $d_z$, and against $X$ errors the
number of rows $d_x$. Throughout, we call the error type against which a
preparation retains its full fault distance the \emph{protected} direction and
the other type the \emph{conjugate} direction.

\subsection{Unitary stabilizer-expanding preparation}\label{sec:background-unitary}
Unitary preparation~\cite{ref:unitaryprep,zen2024rl} initializes qubits in
$\ket{0}$ or $\ket{+}$ and grows the code stabilizers by CNOT layers. For
the surface code, the circuit seeds one line of the patch, and each layer
entangles the next line of fresh qubits, extending every stabilizer on the
growth front until all stabilizers reach full weight. Figure~\ref{fig:construction}(a) shows this
growth directly as stabilizer (detector) slices. Since a CNOT copies $X$
operators from control to target and $Z$ operators from target to control,
the two stabilizer types expand in opposite roles across the same CNOT layer
(top vs.\ bottom row). The crucial ingredient is a careful arrangement of the CNOTs such that
errors propagate only in a correctable way while the stabilizers still
expand~\cite{ref:unitaryprep}, as discussed in Sec.~\ref{sec:method-ft}.

\subsection{Neutral-atom hardware and AOD shuttling}\label{sec:background-na}
Neutral-atom processors hold qubits in static optical-tweezer (SLM) traps and
move selected atoms with AOD tweezers. The native operations are global and
local single-qubit rotations and CZ gates, applied by a single Rydberg pulse
to all pairs within blockade
radius~\cite{levine2019parallel,evered2023gates,bluvstein2024logical}. AOD moves act on whole
rows or columns collectively, must preserve the trap order, and cannot
cross~\cite{bluvstein2022transport,schmid2024na,tan2024compiling}. Architectures use either a
single zone or separate storage, entangling, and readout
zones~\cite{bluvstein2022transport}. Since measurement is orders of
magnitude slower than gates~\cite{graham2023midcircuit}, measurement-free
preparation is attractive, especially on single-zone machines with no readout
zone at all.

\section{Fast Bidirectional Preparation}\label{sec:method}

Recently, a unitary surface-code preparation was
proposed~\cite{ref:unitaryprep}, which we call \emph{single-pivot}. It grows
the patch from a single corner qubit, where attaching each new line costs
three sequential CNOT layers, so the full patch takes $3(d{-}1)$ layers. A first improvement is to exploit
parallelism. Once a qubit has been entangled, it can itself drive the CNOTs
of the next layer, so the qubits entangled in one layer expand the patch
together in the next. The resulting \emph{unidirectional} cascade attaches
one fresh line per CNOT layer and sweeps across the patch in $d$ layers.
Both constructions use the same CNOT arrangement as~\cite{ref:unitaryprep}
and are therefore fault-tolerant in the protected direction.

\begin{figure*}[t]
  \centering
  \vspace{-15pt}
  \begin{minipage}[b]{0.58\textwidth}
    \centering
    \IfFileExists{figures/stab_slices.pdf}
      {\includegraphics[width=\linewidth]{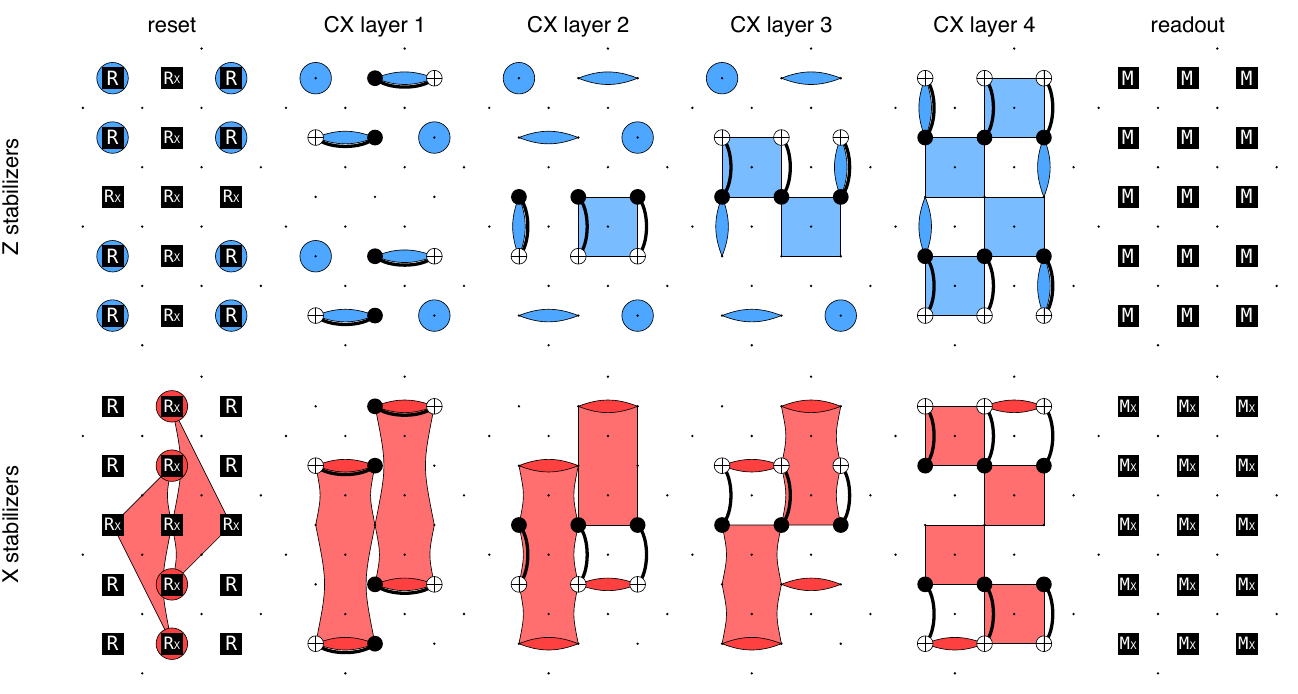}}
      {\fbox{\parbox[c][3cm][c]{0.9\linewidth}{\centering\itshape
        [figures/stab\_slices.pdf pending]}}}
    \par\smallskip {\small (a)}
  \end{minipage}\hfill
  \begin{minipage}[b]{0.32\textwidth}
    \centering
    \includegraphics[width=\linewidth]{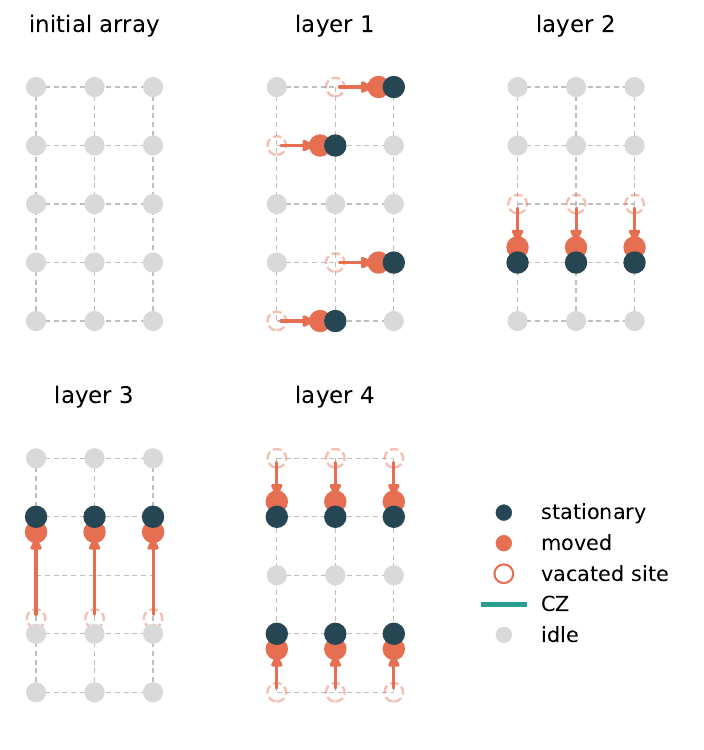}
    \par\smallskip {\small (b)}
  \end{minipage}
  \caption{Bidirectional preparation of a $5\times3$ $\plusl$ patch.
  (a)~Construction as stim detector slices, where the $Z$-type (top, blue) and
  $X$-type (bottom, red) stabilizers grow with each CNOT layer.
  (b)~The same cascade as AOD moves, one panel per layer. Each layer is a
  single collective nearest-neighbour shift followed by a global-Rydberg CZ.}
  \label{fig:construction}
\end{figure*}

But nothing forces the sweep to start at a boundary. Our
\emph{bidirectional} cascade seeds the \emph{middle} line of the patch and
expands in both directions simultaneously. After the two innermost
attachments, which share the seed line and therefore run sequentially, each
layer attaches one fresh line on each side. This halves the depth to
$(d{+}3)/2 = \bigo{d/2}$ without extra qubits and with the same
protected-direction distance, since each half of the patch is a
unidirectional cascade. Figure~\ref{fig:construction} shows a $5\times3$
patch completing in $(5{+}3)/2=4$ CNOT layers, and Table~\ref{tab:schedule}
compares the scaling of the methods.

The construction extends directly to rectangular patches, whose depth
follows the swept dimension only. Since the preparation protects one error
type anyway, asymmetric patches giving more distance to the dominant type
are a natural fit under biased noise~\cite{bonilla2021xzzx}.

\begin{table}[t]
  \centering
  \caption{Schedule scaling of the preparation methods.}
  \label{tab:schedule}
  \begin{tabular}{@{}lccc@{}}
    \toprule
    Construction & CZ layers & AOD shuttles & Row pickups \\
    \midrule
    Single-pivot~\cite{ref:unitaryprep} & $3(d{-}1)$  & $4(d{-}1)$                  & $d{-}1$ \\
    Unidirectional                      & $d$         & $\approx\frac{3}{2}d$       & $\approx\frac{1}{2}d$ \\
    Bidirectional (this work)           & $(d{+}3)/2$ & $\approx\frac{3}{4}(d{+}3)$ & $\approx\frac{1}{4}(d{+}3)$ \\
    \bottomrule
  \end{tabular}
\end{table}

\subsection{Fault tolerance and the conjugate direction}\label{sec:method-ft}
\begin{figure*}[t]
  \centering
  \includegraphics[width=0.68\textwidth]{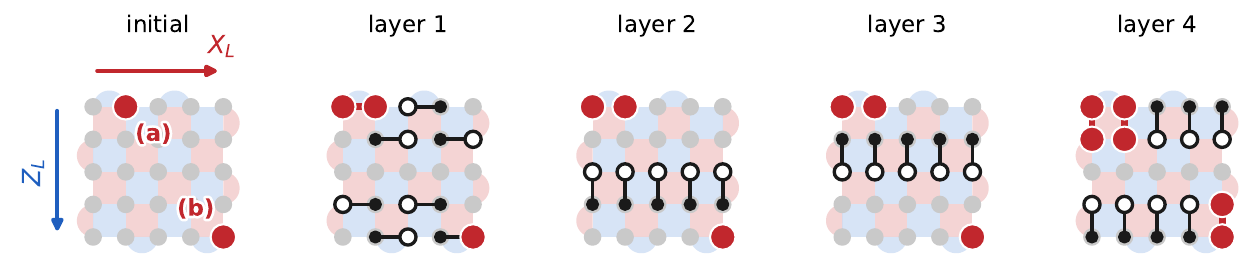}\\[2pt]
  \includegraphics[width=0.68\textwidth]{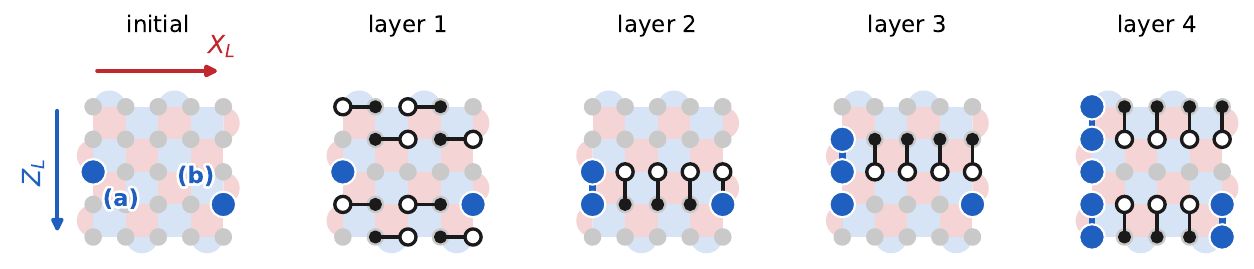}
  \caption{Single-fault propagation through the cascade of a $\zerol$ patch,
  one panel per CNOT layer (propagating CNOTs highlighted).
  \textbf{Top ($X$, protected):} a fault ends as (a)~a stabilizer or (b)~a
  correctable weight-2 error orthogonal to the logical.
  \textbf{Bottom ($Z$, conjugate):} a fault spreads along $Z_L$ to (a)~the
  full logical operator, acting trivially (harmless), or (b)~an irreducible
  weight-2 error (not fault tolerant).}
  \label{fig:errorprop}
\end{figure*}

The fault tolerance of these stabilizer-expanding cascades has been established
in~\cite{ref:unitaryprep}, and since each half of the bidirectional cascade
is such a cascade, the arguments carry over to our construction. Why the
protected direction keeps its full distance is visible directly in the error
propagation through the cascade (Fig.~\ref{fig:errorprop}, top, for a
$\zerol$ patch). A single $X$ fault copies along the CNOTs of subsequent
layers. Because fresh qubits are only targets, its final support, reduced
modulo the code stabilizers, is either trivial (a stabilizer as in
Fig.~\ref{fig:errorprop}(top,~a)), a weight-1 error, or a low-weight error
supported \emph{orthogonally} to the logical string as in
Fig.~\ref{fig:errorprop}(top,~b). All of these are correctable, so the fault
distance in the protected direction remains $d$~\cite{ref:unitaryprep}.

The conjugate type behaves differently (Fig.~\ref{fig:errorprop}, bottom).
$Z$ faults copy in the opposite direction and spread \emph{along} the logical
$Z_L$. A fault on the seed line grows into the full logical as in
Fig.~\ref{fig:errorprop}(bottom,~a), which acts trivially on $\zerol$ and is
therefore harmless. A fault on a later line, however, can leave an
irreducible weight-2 error along the logical as in
Fig.~\ref{fig:errorprop}(bottom,~b), the same error propagation structure
identified in~\cite{ref:unitaryprep}. The conjugate direction is therefore not
fault-tolerant. For a stored $\zerol$ this is benign, since conjugate errors
commute with the state and its readout, but it precludes using the fresh
patch directly as a Steane-type ancilla~\cite{steane1997active} for a
transversal CX~\cite{bluvstein2024logical} without an intermediate QEC round
after the state preparation.

\section{Neutral-Atom Realization}\label{sec:na-realization}

\subsection{Mapping the cascade to atom moves}\label{sec:na-moves}
We now discuss how the bidirectional cascade maps efficiently onto
neutral-atom hardware. Each CNOT layer compiles to single-qubit basis changes
plus one CZ layer, executed by a single global Rydberg pulse over all paired
atoms~\cite{evered2023gates}. Bringing the pairs together requires exactly
one collective nearest-neighbour line shift per
layer~\cite{bluvstein2022transport}. The moving line is displaced by one
lattice step onto its partner line, entangled, and later returned
(Fig.~\ref{fig:construction}(b)). All moves are parallel shifts of whole
lines, so they preserve the trap order and never cross, fulfilling the AOD
constraints.

The shuttling cost amortizes further. The first layer costs one row pickup
and two shuttles (onto the partner and back). Afterwards, a picked row serves
two consecutive CZ layers by moving onto its first partner, directly on to
its second, and only then back, which amounts to three shuttles and one
pickup per two layers, as visible in Fig.~\ref{fig:construction}(b). For the
bidirectional cascade this yields $\approx\frac{3}{4}(d{+}3)$ collective
shuttles and $\approx\frac{1}{4}(d{+}3)$ pickup--drop cycles, so global
pulses, shuttles, and pickups all scale as $\bigo{d/2}$. The unidirectional sweep
follows the same accounting at twice the layer count, and the single-pivot
construction needs four shuttles and one pickup per attached line
(Table~\ref{tab:schedule}).

\subsection{Hardware noise model}\label{sec:na-noise}
We evaluate the schedule under the hardware-calibrated, circuit-level noise
model of the open-source \texttt{bloqade} SDK~\cite{tool:bloqade}, which
compiles the circuit to the native gate set and attaches Pauli channels to
every gate, move, and idle period. Table~\ref{tab:noise} lists the key
rates. The dominant channels (CZ and shuttling) are strongly $Z$-biased, and
the CZ additionally carries a correlated two-qubit channel dominated by its
$IZ/ZI/ZZ$ terms.

\begin{table}[t]
  \centering
  \caption{Key Pauli rates of the neutral-atom noise model~\cite{tool:bloqade}.}
  \label{tab:noise}
  \begin{tabular}{@{}lcc@{}}
    \toprule
    Channel (per qubit) & $p_X = p_Y$ & $p_Z$ \\
    \midrule
    CZ, paired atoms    & $6.5\times10^{-4}$ & $3.2\times10^{-3}$ \\
    CZ, unpaired atoms  & $5.1\times10^{-4}$ & $2.2\times10^{-3}$ \\
    Move (AOD shuttle)  & $8.1\times10^{-4}$ & $2.5\times10^{-3}$ \\
    Idle                & $3.1\times10^{-4}$ & $4.6\times10^{-4}$ \\
    1q gate, local      & $4.1\times10^{-4}$ & $4.1\times10^{-4}$ \\
    \bottomrule
  \end{tabular}
\end{table}

\begin{figure*}[t]
  \centering
  \vspace{-24pt}%
  \includegraphics[width=0.69\textwidth]{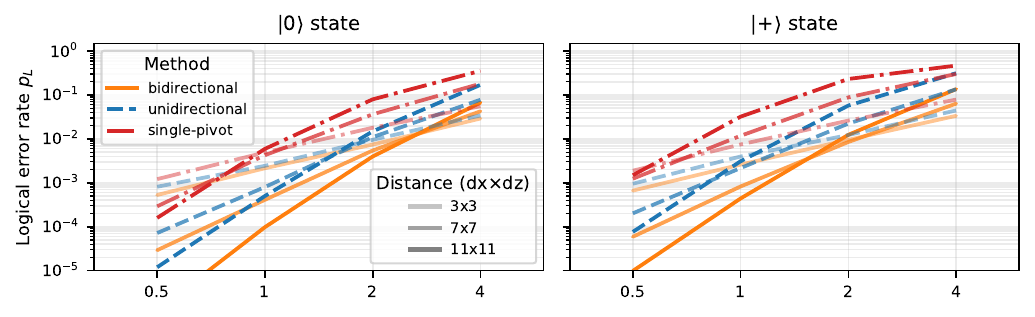}
  \vspace{-6pt}%
  \caption{Logical error rate vs.\ noise scaling factor ($1=$ device-level
  neutral-atom noise). Colour and line style denote the construction, line
  opacity the distance.}
  \label{fig:scaling}
\end{figure*}

\section{Evaluation}\label{sec:eval}

To quantify the benefit of the shorter schedules, we compare all three
constructions for both $\zerol$ and $\plusl$ at distances 3, 7, and 11,
sweeping a global scaling factor on all noise rates of
Table~\ref{tab:noise}. Each experiment prepares the logical state, measures
all data qubits in the matching basis, and decodes the reconstructed
stabilizers with minimum-weight perfect matching
(Stim~\cite{tool:stim} $+$ PyMatching~\cite{tool:pymatching}), with up to
$10^7$ shots per point. Code and evaluation data are available at
\url{https://github.com/lsschmid/fast-surface-code}.

Figure~\ref{fig:scaling} shows the logical error rates for both states. At every distance and noise scale, the constructions order as
bidirectional $<$ unidirectional $<$ single-pivot. This ordering reflects
the circuit volume, since fewer layers mean fewer global
pulses, shuttles, and idle periods, and therefore less exposure to the
dominant noise channels of Table~\ref{tab:noise}.

We further observe that $\zerol$ reaches substantially lower error rates
than $\plusl$, which follows from the noise bias. The dominant channels are
$Z$-biased, and $Z$ errors are harmless for $\zerol$, since $Z_L$ stabilizes
the state and the $Z$-basis readout is blind to them, whereas for $\plusl$
they are exactly the dangerous error type.

Finally, at the considered device-level noise (scaling factor $1$), only
the bidirectional construction is below threshold for both states, meaning
its logical error rate falls with increasing distance. The unidirectional
cascade achieves this only for $\zerol$, the single-pivot for neither. This
shows that, at current hardware error rates, optimizing
the state preparation for neutral atoms is essential for sub-threshold
behaviour.

\section{Discussion and Conclusion}\label{sec:discussion}

In this work, we improved the unitary, measurement-free preparation of
surface-code logical states of~\cite{ref:unitaryprep} and adapted it to
neutral-atom hardware and its AOD constraints. Growing the patch
bidirectionally from its middle line halves the preparation depth and maps
onto $\bigo{d/2}$ global Rydberg pulses, collective nearest-neighbour
shuttles, and row pickups, without any measurement or transport to a readout
zone. With the open-source neutral-atom toolchain
\texttt{bloqade}, we compiled the schedules to the native gate set and
evaluated them under hardware-calibrated device noise. The proposed
construction outperforms the original single-pivot proposal
of~\cite{ref:unitaryprep} and the unidirectional sweep at every distance,
and is the only construction tested that operates below threshold for both
logical states at the considered device-level noise, making it a practical
candidate for logical-state preparation in experiments.
The preparation is fault-tolerant against one error type, while the
conjugate type spreads along the logical but stabilizes the prepared state.
A single subsequent QEC round restores fault tolerance against both
types~\cite{ref:unitaryprep}.

\section*{Acknowledgment}
\begingroup\scriptsize\linespread{0.95}\selectfont
L.S.\ would like to thank Tom Peham and Katharina K\"ostler for helpful
discussions.
The authors acknowledge funding from the European Research Council (ERC) under
the European Union's Horizon 2020 research and innovation program grant
agreement No.~101001318 and No.~101114305, and
the Munich Quantum Valley (MQV), which is supported by the Bavarian
state government with funds from the Hightech Agenda Bayern Plus. Furthermore, this work was supported by the Deutsche Forschungsgemeinschaft (DFG, German Research Foundation) under grant numbers 563402549.
The software used to generate the figures and evaluation data in this work, and
initial drafts of the manuscript text, were prepared with the assistance of a
generative AI system (Anthropic Claude~\cite{ai:claude}); all code and results
were verified, and the manuscript reworked and finalized, by the authors, who
take responsibility for all content.\par
\endgroup

\bibliographystyle{IEEEtran}
\begingroup\linespread{0.93}\selectfont
\bibliography{references}
\endgroup

\end{document}